# Time- and angle-resolved photoemission spectroscopy with wavelength-tunable pump and extreme ultraviolet probe enabled by twin synchronized amplifiers

Takeshi Suzuki[1*], Yigui Zhong[1], Kecheng Liu[1], Teruto Kanai[1], Jiro Itatani[1], and Kozo Okazaki[1,2,3†]

[1]*The Institute for Solid State Physics, The University of Tokyo, Kashiwa, Chiba 277-8581 Japan*
[2]*Material Innovation Research Center, The University of Tokyo, Kashiwa, Chiba 277-8561, Japan*
[3]*Trans-scale Quantum Science Institute, The University of Tokyo, Bunkyo-ku, Tokyo 113-0033, Japan*

Abstract

We describe a setup for time- and angle-resolved photoemission spectroscopy with wavelength-tunable excitation and extreme ultraviolet probe. It is enabled by using the 10 kHz twin Ti:sapphire amplifiers seeded by the common Ti:sapphire oscillator. The typical probe energy is 21.7 eV, and the wavelength of the pump excitation is tuned between 2400 and 1200 nm by using the optical parametric amplifier. The total energy resolution of 133 meV is achieved, and the time resolution is dependent on the wavelength for the pump, typically better than 100 fs. This system enables the pump energy to be matched with a specific interband transition and to probe a wider energy-momentum space. We present the results for the prototypical materials of highly oriented pyrolytic graphite and $Bi_2Se_3$ to show the performance of our system.

**I.    Introduction**

Angle-resolved photoemission spectroscopy (ARPES) is a very powerful method in solid state physics because it can directly capture the band structure of a material [1]. Integrating the pulsed laser into ARPES can achieve the time-resolved measurements of ARPES, which enables to capture the nonequilibrium phenomena in the field of ARPES [2] [3]. In particular, the mode-locking and amplification techniques using a Ti:sapphire crystal as a gain medium enabled sufficient photon flux to be achieved within an extreme ultraviolet wavelength region by using wavelength conversion techniques, and the time resolution can be achieved at a femtosecond time scale [4] [5] [6] [7] [8]. However, photon energies of ~6 eV are more commonly used as a light source for time- and angle-resolved photoemission spectroscopy (TARPES) through the use of up-conversion techniques with a nonlinear crystal such as β-$BaB_2O_4$(BBO), and the energy and momentum regions accessible with these photon energies are significantly limited [9] [10] [11].

Alternatively, high harmonic generation (HHG) techniques using noble gas overcome this limitation, and can generate much higher-order harmonics within the energy region of 10–70 eV, and enable access to full valence bands, shallow core levels, and a wide momentum space [12] [13] [14] [15] [16] [17] [18] [19] [20] [21] [22]. For example, the atomic site-selective investigation in $TaS_2$ was enabled by accessing core levels, which successfully revealed the electron transfer dynamics between Ta atoms during the excitation of the CDW amplitude mode [23]. Another example is found in graphene possessing Dirac bands at the K point, where relaxation mechanisms of the Dirac electrons were intensively studied [24].

As for the pump, on the other hand, most of the TARPES systems use fixed spectra in the near-infrared region determined by the fundamental light source. For example, the center wavelength of Ti:sapphire lasers is 800 nm, and that of Yb:KGW laser is 1030 nm. While we have successfully revealed many interesting nonequilibrium phenomena in quantum materials using near-infrared pulses for excitation [25] [26] [27] [28] [29] [30] [31], they usually excite the electronic system. Moreover, because TARPES typically targets the metallic or semi-metallic systems with many bands lying around $E_F$, the NIR pulse excites many electron-hole pairs separated by the energy of the pump. Such a non-resonant or non-selective excitation inevitably generates substantial heat in the system, which hinders discovering the fundamental phenomena purely driven by the rearrangement of the electronic distribution. For example, the resonant excitation of the 1s exciton state is known to significantly increase Mott transition density compared to the non-resonant excitation [32] [33]. Furthermore, the excitation of the lattice system can be possible by using mid-infrared light to excite the infrared active phonons, by which very exotic phenomena such as photoinduced superconductivity [34] or ferroelectricity [35] have been reported.

In this paper, we report our development of a new TARPES system by integrating wavelength-tunable excitation in our system. First, we extensively describe our experimental setup and profiles of the pump and probe. Then we discuss the results using a new setup to investigate prototypical samples of highly oriented pyrolytic graphite (HOPG) and $Bi_2Se_3$. Finally, we briefly discuss an outlook for the future progress by using this system.

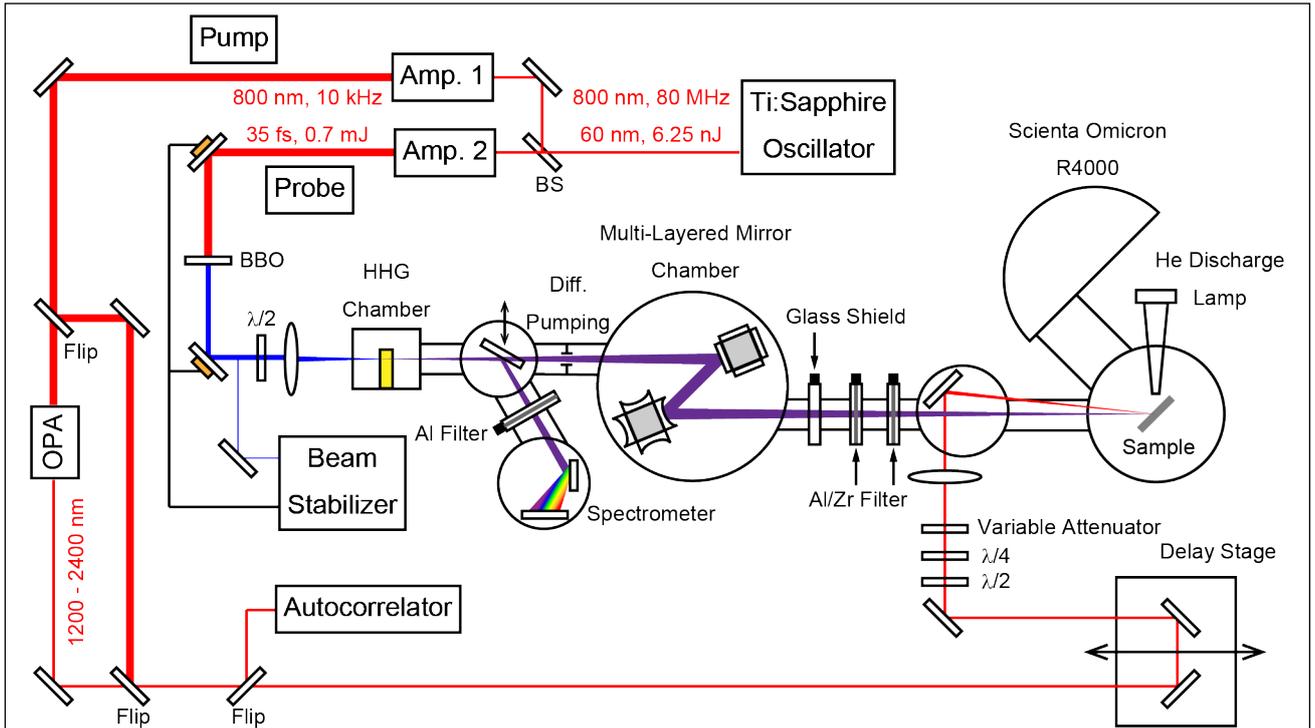

**FIG. 1** Schematic illustration of the twin laser-based time- and angle-resolved photoemission spectroscopy (TARPES) system composed of two Ti:sapphire amplification systems, optical parametric amplifier (OPA), beam stabilizer, high harmonic generation (HHG) chamber, spectrometer, multi-layered mirror chamber, and hemispherical electron analyzer.

## II. Experimental setup
### A. TARPES layout

Figure 1 shows the layout of our TARPES setup. The twin Ti:sapphire amplifiers (Spectra Physics, Solstice Ace) are used. The center wavelength is 800 nm, the time duration is 35 fs, the pulse energy is 0.7 mJ, and the repetition rate is 10 kHz for each amplifier. In our TARPES setup, one amplifier is used for the pump and the other is used for the probe. They are seeded by the common Ti:sapphire oscillator (Spectra Physics, MaiTai). By using the common seed laser, the timing jitter between the two amplifiers is significantly reduced, resulting in no deterioration of the time resolution in our setup. This is extremely powerful compared to the typical setup using the electrical synchronization technique. Another advantage of using the twin amplifiers is that we can gain a total power of higher than 10 W. Typically, to achieve more than 10 W power by a single amplifier, the cryogenic cooling setup for the gain crystal is necessary to avoid heating.

As for the probe, we first double the photon energy to 3.10 eV by generating second harmonic (SH) of fundamental light from a BBO crystal and then focus the pulse on Ar gas filled in the gas cell to achieve HHG. The thickness of the gas cell is ~4 mm and the back pressure of Ar gas is 3-5 Torr. Since the output of

HHG is very sensitive to the focused beam position, we use the beam stabilizer (TEM Messtechnik GmbH, Aligna) that steers the two mirrors forming a feedback loop, which suppresses thermal drifts as well as fast mechanical fluctuations. We typically select the 7th harmonic corresponding to 21.7 eV by using a pair of SiC/Mg multilayer mirrors (MLM). Between the HHG chamber and the MLM chamber, we use the differential pumping, which enables the relatively high pressure of ~$10^{-3}$ Torr in the HHG chamber while maintaining the pressure of the MLM chamber to ~$10^{-8}$ Torr. To block the fundamental light, we use the Al or Zr filters with a thickness of 50 or 100 nm for each. The total photon flux of HHG at the sample position is evaluated as ~$10^9$ photons/s. For evaluating sample quality and orientation, the He discharge lamp (Scienta Omicron, VUV5000) is also equipped to measure photoemission spectra in the equilibrium condition.

As for the pump, we use the optical parametric amplifier (OPA) (LIGHT CONVERSION, TOPAS-Prime) to enable the wavelength-tunable excitation between 1160 and 2580 nm. The autocorrelator (APE GmbH, pulseCheck) is used for monitoring the time profile of the pump. The delay stage is placed on the beam path for the pump to change the delay between the pump and probe. The half ($\lambda/2$) and quarter ($\lambda/4$) waveplates are inserted to change the polarization of the pump. The pump fluence is changed by using the variable attenuator. Also, we have an option to employ fundamental light of 800 nm [36], by activating the

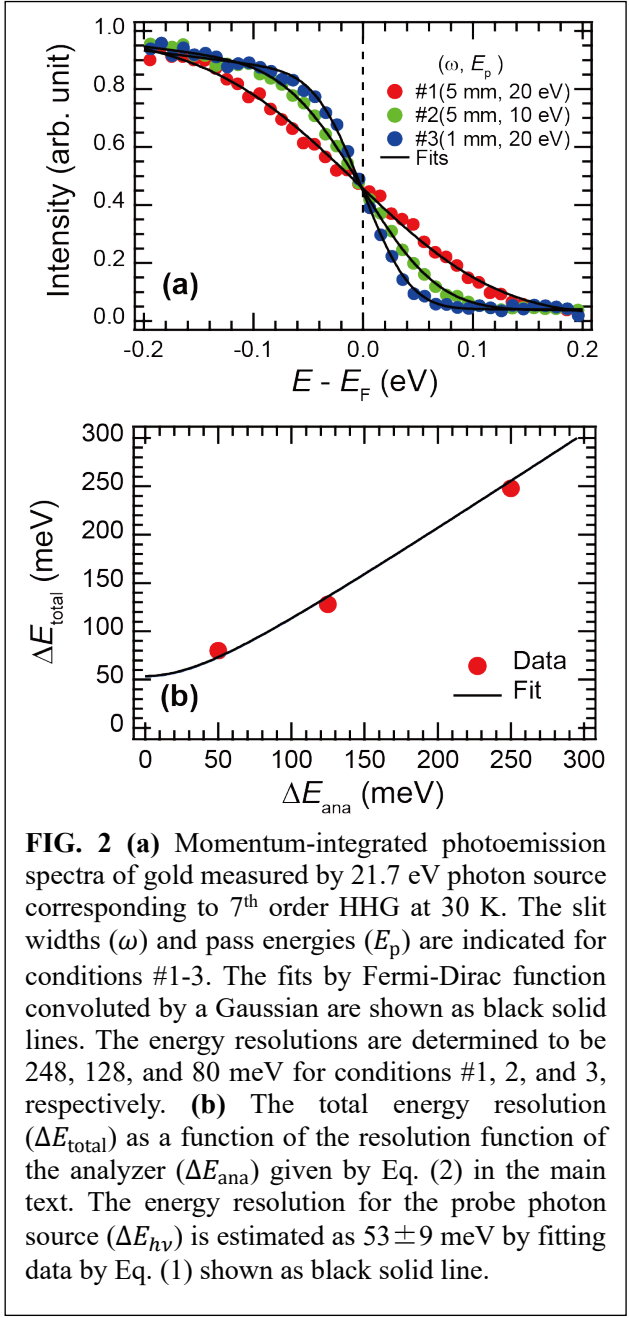

FIG. 2 (a) Momentum-integrated photoemission spectra of gold measured by 21.7 eV photon source corresponding to 7$^{th}$ order HHG at 30 K. The slit widths ($\omega$) and pass energies ($E_p$) are indicated for conditions #1-3. The fits by Fermi-Dirac function convoluted by a Gaussian are shown as black solid lines. The energy resolutions are determined to be 248, 128, and 80 meV for conditions #1, 2, and 3, respectively. (b) The total energy resolution ($\Delta E_{\text{total}}$) as a function of the resolution function of the analyzer ($\Delta E_{\text{ana}}$) given by Eq. (2) in the main text. The energy resolution for the probe photon source ($\Delta E_{h\nu}$) is estimated as 53 ± 9 meV by fitting data by Eq. (1) shown as black solid line.

analyzer. Conditions #1, 2, and 3 correspond to the (slit width ($\omega$), pass energy ($E_p$)) of (5 mm, 20 eV), (5 mm, 10 eV), and (1 mm, 20 eV), respectively. The measured spectra are fitted by the Fermi-Dirac function convoluted with a Gaussian shown as black solid lines. The energy resolutions for the conditions #1-3 corresponding to the full width of the half maximum (FWHM) of the Gaussian are 248, 128, and

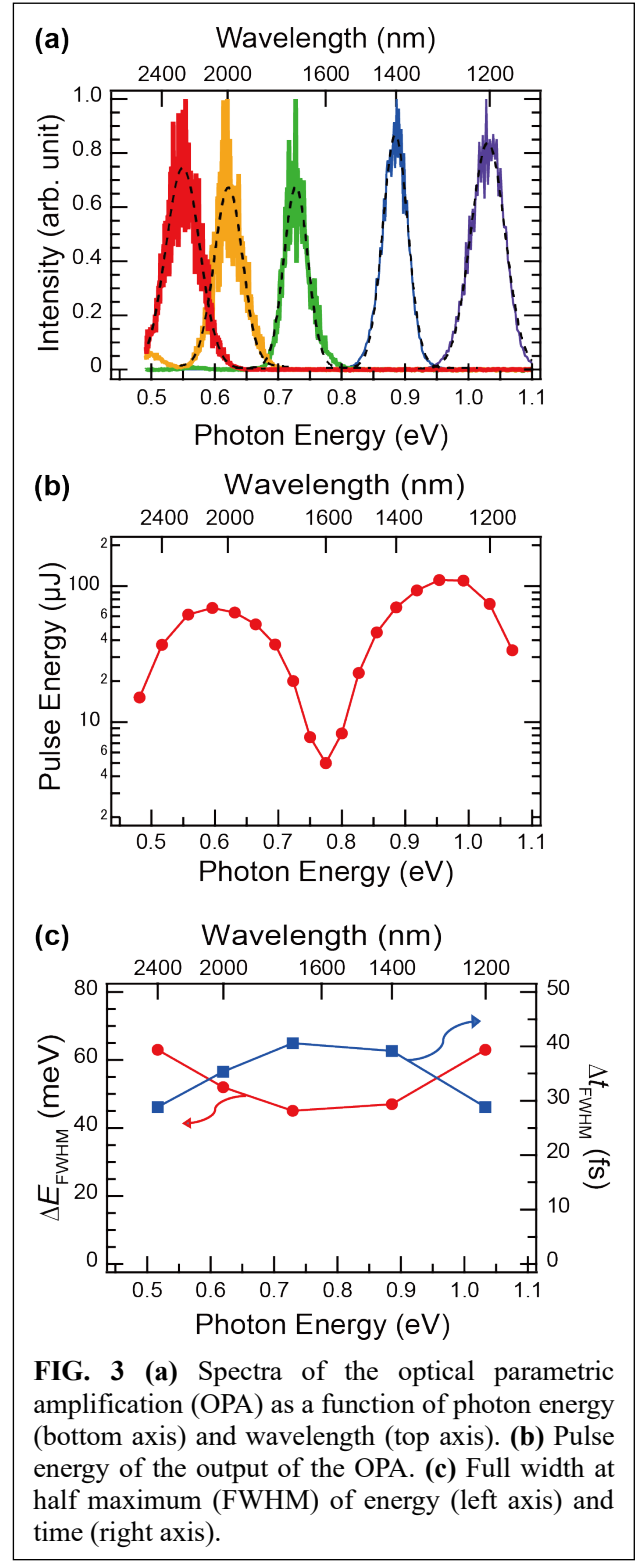

FIG. 3 (a) Spectra of the optical parametric amplification (OPA) as a function of photon energy (bottom axis) and wavelength (top axis). (b) Pulse energy of the output of the OPA. (c) Full width at half maximum (FWHM) of energy (left axis) and time (right axis).

flipping mirror.

The photoelectrons are collected using a hemispherical electron analyzer (Scienta Omicron, R4000). The base pressure of the analyzer chamber is ~2×10$^{-11}$ Torr. The cryostat is also equipped to allow temperature to cool down to 7 K.

**B. HHG spectra**

To evaluate the spectral resolution, we measure the Fermi edge broadening of polycrystalline Au films, which are evaporated to the tip of the sample holder in vacuum. Figure 2 shows the momentum-integrated photoemission spectra of Au measured by 21.7 eV photons obtained by HHG. The temperature is set to 30 K. The measurements are performed using three representative conditions of the hemispherical electron

80 meV, while the photoemission count rates at the Fermi level for #1-3 are 4,000, 2,000, and 1,000 counts/s. We can use the condition #3 for the samples emitting sufficiently large amounts of photoelectrons. To avoid the space charge effect, which gives rise to the undesirable shift or broadening of the photoemission spectra due to Coulomb repulsion of photoelectrons, we must reduce photoemission count rate. The space charge effect depends on the sample and surface quality, and thus we must check every time for the measurements. In case of Au, the space charge effect gives a serious problem when the photoemission count rate is larger than 2,000 counts/s for the measurement condition #3.

There are several factors determining the total energy resolution ($\Delta E_{\text{total}}$) for the photoemission spectroscopy. Mainly, the probe photon source ($\Delta E_{h\nu}$) and the analyzer ($\Delta E_{\text{ana}}$) contribute to the total energy resolution, as given by

$$\Delta E_{\text{total}} = \sqrt{(\Delta E_{h\nu})^2 + (\Delta E_{\text{ana}})^2}. \quad (1)$$

The energy resolution of the analyzer is given by

$$\Delta E_{\text{ana}} = E_p \frac{\omega}{2R}, \quad (2)$$

where $R$ is the radius of the hemispherical analyzer, and $R = 200$ mm for Scienta Omicron R4000. Using Eq. (2), $\Delta E_{\text{ana}}$ is estimated as 250, 125, and 50 meV for the conditions #1, 2, and 3, respectively. We plot the total energy resolution measured by the experiment for #1-3 in Fig 2(b) as a function of $\Delta E_{\text{ana}}$. To evaluate $\Delta E_{h\nu}$, we fit our data by Eq. (1) shown as black solid line, and $\Delta E_{h\nu}$ is estimated to be $53 \pm 9$ meV. The minimum value of the time duration of the probe pulse is estimated to be 34 fs in the Fourier limit from the spectral width.

## C. OPA pump spectra and time duration

For the pump, we employ the OPA to enable the wavelength-tunable excitation. As we mentioned, the center wavelength of the fundamental light source is 800 nm with a time duration of 35 fs and pulse energy of 0.7 mJ. Figure 3(a) shows the spectra of the output of the OPA for representative center wavelengths. The ranges of the photon energy (wavelength) for the signal and idler correspond to [1.07 eV, 0.77 eV] ([1160 nm, 1600 nm]) and [0.77 eV, 0.48 eV] ([1600 nm, 2580 nm]), respectively. Figure 3(b) shows the pulse energy as a function of the photon energy (wavelength). While the fluences are significantly dependent on wavelength, they exceed 5.0 mJ/cm² at any wavelength at the sample position after focusing.

To measure the spectral widths of each wavelength, we fit them by the Gaussians shown as black dashed lines in Fig. 3(a). The measured widths (FWHM) of the spectra are shown in Fig. 3(c) as red-solid circles as a function of photon energy (wavelength). The time duration is evaluated by assuming a time-bandwidth product of a Gaussian profile and found to be shorter than 45 fs for all the wavelengths. The results are shown as blue solid squares in Fig. 3(c).

To directly measure the time duration of the output of the OPA, we perform the autocorrelation (AC) measurements. Figure 4(a) shows the AC signals for the output of the OPA, whose center photon energies (wavelengths) are 0.52 (2400) and 1.03 eV (1200 nm). As clearly seen, the 0.52 eV pulse has a much broader temporal profile than the 1.03 eV pulse. The temporal widths are estimated by fitting the AC signals by the Gaussians shown as black solid lines. Figure 4(b) shows the summary of the time duration (FWHM) of the pump ($\Delta t_{\text{pump}}$) as a function of the photon energy (wavelength). Fourier-limited values from the spectral width (Fig. 3(c)) are also shown. It is clearly found that the measured time durations shown in Fig. 4(b) have longer values than the Fourier-limited values from the

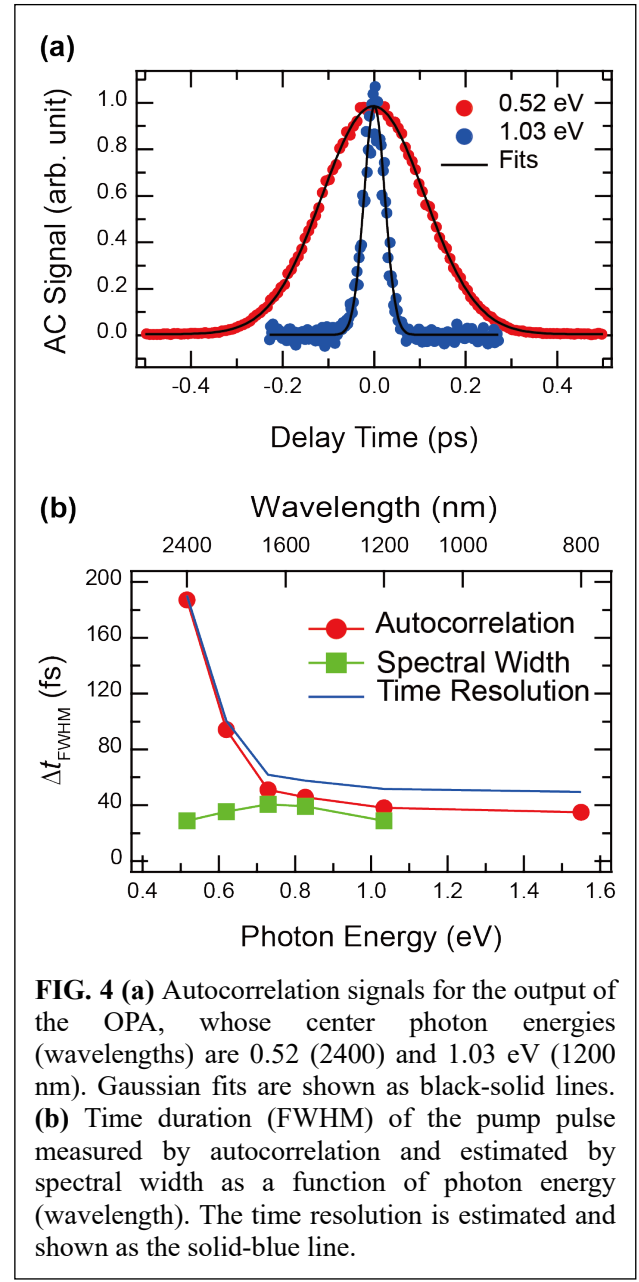

FIG. 4 (a) Autocorrelation signals for the output of the OPA, whose center photon energies (wavelengths) are 0.52 (2400) and 1.03 eV (1200 nm). Gaussian fits are shown as black-solid lines. (b) Time duration (FWHM) of the pump pulse measured by autocorrelation and estimated by spectral width as a function of photon energy (wavelength). The time resolution is estimated and shown as the solid-blue line.

spectral width. Especially, the longer wavelength has a longer time duration. This trend is especially significant for the pulses whose center photon energies (wavelengths) are smaller (longer) than 0.62 eV (2000 nm). This is due to the chirp caused by the nonlinear crystals inside the OPA system, which can be in principle improved by pulse compression using *e.g.*, chirp mirrors, gratings, or prisms. However, the pulse compression also accompanies the different optical paths for each wavelength and unavoidably degrades the pulse profile, which makes the wavelength-dependent measurements for the same cleaved sample unpractical.

Furthermore, assuming the gaussian profile in time domain for pump and probe, the time resolution ($\Delta t_{res}$) is given by $\Delta t_{res} = \sqrt{(\Delta t_{pump})^2 + (\Delta t_{probe})^2}$, where $\Delta t_{probe}$ is the time duration of the probe. Assuming the Fourier limit with a reasonable error, $\Delta t_{probe}$ is determined to be $35 \pm 10$ fs. $\Delta t_{res}$ is also shown in Fig. 4(b) with $\Delta t_{probe} = 35$ fs and it is clearly seen that $\Delta t_{res}$ is mostly determined by $\Delta t_{pump}$ below 0.8 eV. Then, the response time of the sample ($\Delta t_{sample}$) is extracted from the experimentally measured time duration ($\Delta t_{meas}$) by using the relationship of

$$\Delta t_{meas} = \sqrt{(\Delta t_{pump})^2 + (\Delta t_{probe})^2 + (\Delta t_{sample})^2}. \quad (3)$$

### D. Spatial overlap and beam profiles for the pump and probe

To achieve the spatial overlap between the pump and probe beams, we first place the gold at the focus position for the electron analyzer. This is achieved by maximizing the photoemission counts generated by the probe pulse by adjusting the sample position along the normal direction to the detector. Then we insert a glass shield and take out an Al/Zr filter to pass the SH. Its position can be monitored by the CCD camera placed behind the electron analyzer.

The pump beam of OPA is accompanied by the visible light that is generated in the white light generation process in the OPA and the power is significantly small. Thus, the visible light can be used for the beam guide of the pump beam and blocked by Si during the TARPES measurements. We align the pump beam at the same position of the gold as the probe beam by monitoring the CCD camera and thus spatial overlap is achieved.

For the beam size, the diameter of the probe beam was measured to be 200 μm as a FWHM of the power. On the other hand, that of the pump beam can be adjusted by changing the lens position before the chamber. The minimum size is typically 400 μm. For the pump and probe experiments, generally, the larger beam size of the pump is ideal because it can achieve the homogeneous excitation along the in-plane direction.

## III. Performance test

We test our machine by measuring HOPG and bulk $Bi_2Se_3$ crystals, which are prototypical samples for TARPES measurements. We fixed the probe energy to 21.7 eV and changed the photon energies (wavelengths) of the pump between 0.52 (2400) and 1.03 eV (1200 nm). As a reference, we also show the result using an 800 nm pump. In both materials, the pump fluence is fixed to be 1.0 mJ/cm$^2$ for each photon energy, and each data set with a single pump energy and 51 delay steps was accumulated over one day (~ 24 hours).

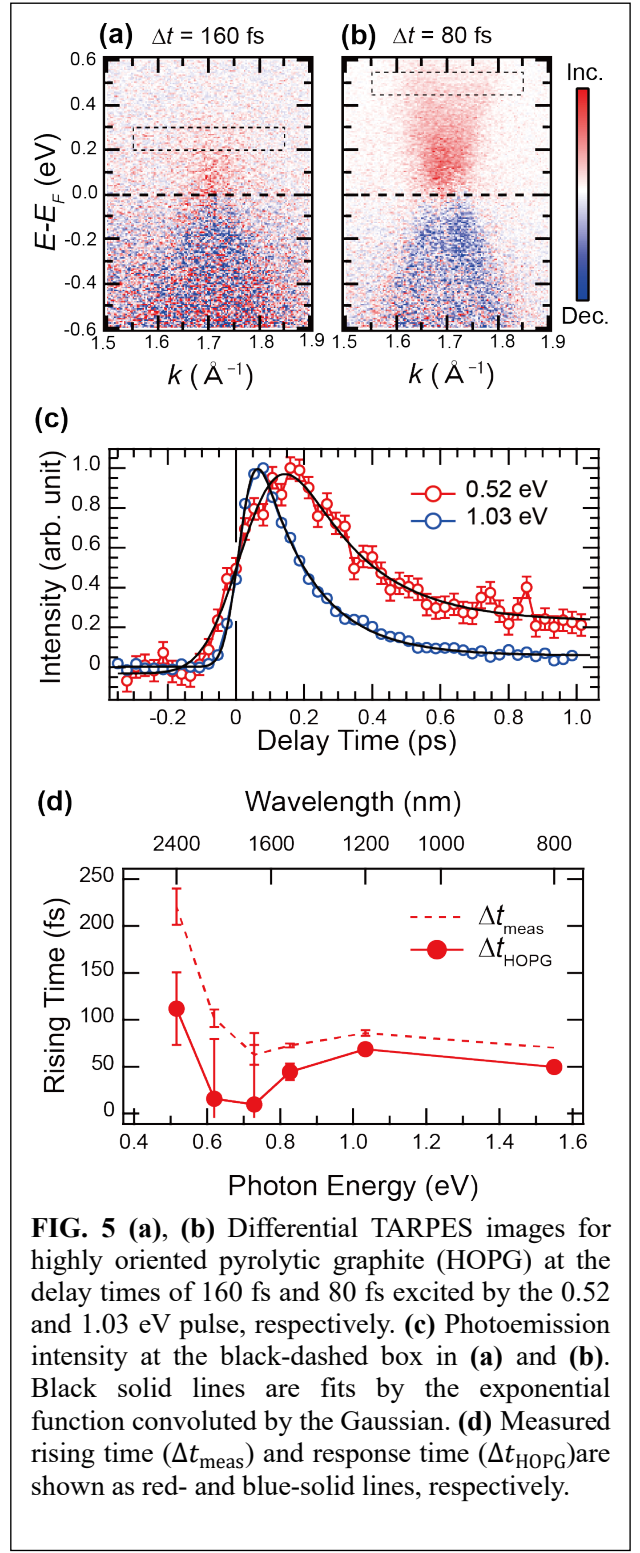

FIG. 5 **(a)**, **(b)** Differential TARPES images for highly oriented pyrolytic graphite (HOPG) at the delay times of 160 fs and 80 fs excited by the 0.52 and 1.03 eV pulse, respectively. **(c)** Photoemission intensity at the black-dashed box in **(a)** and **(b)**. Black solid lines are fits by the exponential function convoluted by the Gaussian. **(d)** Measured rising time ($\Delta t_{meas}$) and response time ($\Delta t_{HOPG}$) are shown as red- and blue-solid lines, respectively.

## A. HOPG

Graphite possesses unique electronic and lattice profiles and has been intensively studied for decades [37]. The hallmark is the Dirac band lying at the K point in the momentum space, which gives rise to many unique properties owing to the Dirac fermion. Furthermore, graphite is closely related to carbon-based nanomaterials including graphene, which is the focus of the ongoing study in nanotechnology. For the study of ultrafast carrier relaxation dynamics, TARPES measurements have also been performed on HOPG many times [38] [39] [40].

Figure 5 shows the summary of the TARPES results on HOPG. The Dirac point ($E_D$) is located at the Fermi level ($E_F$). We use five different photon energies for the pump. Figures 5(a) and 5(b) show the differential TARPES images excited by 0.52 and 1.03 eV pulses, respectively, by subtracting the image before arrival of pump. The delay time is set to be the largest change, corresponding to 160 and 80 fs for Figs. 5(a) and 5 (b), respectively. Since the Dirac band has the linear dispersion symmetric about the Dirac point ($E = E_D$), the electrons are excited from $E_D - 1/2\hbar\omega$ to $E_D + 1/2\hbar\omega$ by pump with energy $\hbar\omega$. To track the initial dynamics after photoexcitation, we highlight the photoexcited electrons at $E = E_D + 1/2\hbar\omega$, which correspond to the $E - E_F = 0.25$ and 0.5 eV for 0.52 and 1.03 eV pumps, respectively. Figure 5(c) shows the time-dependent photoemission intensity integrated within the black dashed boxes in Figs. 5(a) and 5(b). Each box corresponds to $E = [0.2, 0.3]$ eV and $k = [1.55, 1.85]$ Å$^{-1}$ for 0.52 nm (Fig. 5(a)), and $E = [0.45, 0.55]$ eV and $k = [1.55, 1.85]$ Å$^{-1}$ for 1.03 eV (Fig. 5(b)), respectively. The dynamics of the 0.52 eV pump show a slow rising time compared to that of the 1.03 eV pump. To extract rising time quantitatively, we fit by the exponential functions convoluted by the Gaussians, shown as the black-solid lines in Fig. 5(c). The rising times corresponding to FWHM of the fitted Gaussians are shown as a function of photon energy (wavelength) in Fig 5(d), which is referred to as $\Delta t_\text{meas}$. We also show the response time of HOPG ($\Delta t_\text{HOPG}$) using Eq. (3). The non-monotonic behavior having the minimum value at around 0.7 eV might be due to the electron-phonon scattering because the initial electron dynamics are also affected by the phonons with specific energies and momentum, which result in the energy-dependent electron redistribution [41].

## B. Bi$_2$Se$_3$

Bi$_2$Se$_3$ is a representative topological insulator, where the striking feature is found at the topological surface state that has a linear band dispersion with chiral spin texture. Owing to the unique electron and spin properties at the surface state, spin-conserving electrons' scattering events are significantly reduced, which gives rise to the potential of the platform for future spintronic and electronic devices. For investigating the electrons' scattering events in the ultrafast time scale, many TARPES measurements have been reported [42] [43] [44] [45].

Figure 6 summarizes the TARPES results on Bi$_2$Se$_3$. Figures 6(a) and 6(i) show the static ARPES measured by using He discharge lamp. It is found that Bi$_2$Se$_3$ is slightly $n$-doped, and $E_F$ lies at the conduction band. Due to the relationship between $k_z$ and the incident photon energy (kinetic energy of the emitted photoelectrons), it is known that photon energy around 20 eV probes around the Γ point and the valence and conduction bands are located near $E_F$ and appear very clearly [46] [47]. The Dirac point is located at $E - E_F = -0.3$ eV, and the surface Dirac band weakly appears in Figs. 6(a) and 6(i). The black arrows in Figs. 6(a) and 6(i) represent the optical excitations by the pump photon energies of 0.52 and 1.03 eV, respectively. Figures 6(b)-6(e) show the TARPES images of Bi$_2$Se$_3$ at the delay time denoted in each image for the pump photon energy of 0.52 eV. To make the photo-induced changes clearer, the differential images are also shown in Figs. 6(f)-6(h) by subtracting the image before the pump arrival from each image. Figures 6(j)-6(m) and 6(n)-6(p) are TARPES and differential TARPES images, respectively, for the pump photon energy of 1.03 eV.

We first focus on the dynamics at the bulk conduction band. Figure 6(q) shows the time-dependent photoemission intensity integrated within the solid-box in Figs. 6(f) and 6(n) corresponding to $E - E_F = [0, 0.1]$ eV and $k = [-0.1, 0.1]$ Å$^{-1}$. We also show the result of 0.73 eV. We see that the initial rising time for 0.73 eV is faster than those for 0.52 and 1.03 eV while the decay time are similar between them. To see the initial dynamics in more detail, we fit the data by the single exponential decay function convoluted with a gaussian. The rising time and response time by using Eq. (3) are shown as the red-dashed and red-solid lines, respectively, in Fig. 6(s) as a function of pump photon energy. Overall, the response time of the conduction band increases with the pump photon energy with the minimized value at the 0.73 eV.

We next investigate the dynamics below $E_F$. We see that the significant increase of photoelectrons below $E_F$ although it is not trivial because the electrons are not transferred by photo excitation below $E_F$ due to Pauli blocking. One possible reason is the creation of the replica bands induced by laser-assisted photoemission (LAPE) [48] [49] [50] or the Floquet mechanism [51]. In either case, the replica band is seen when the pump and probe are overlapped in time domain, and the time

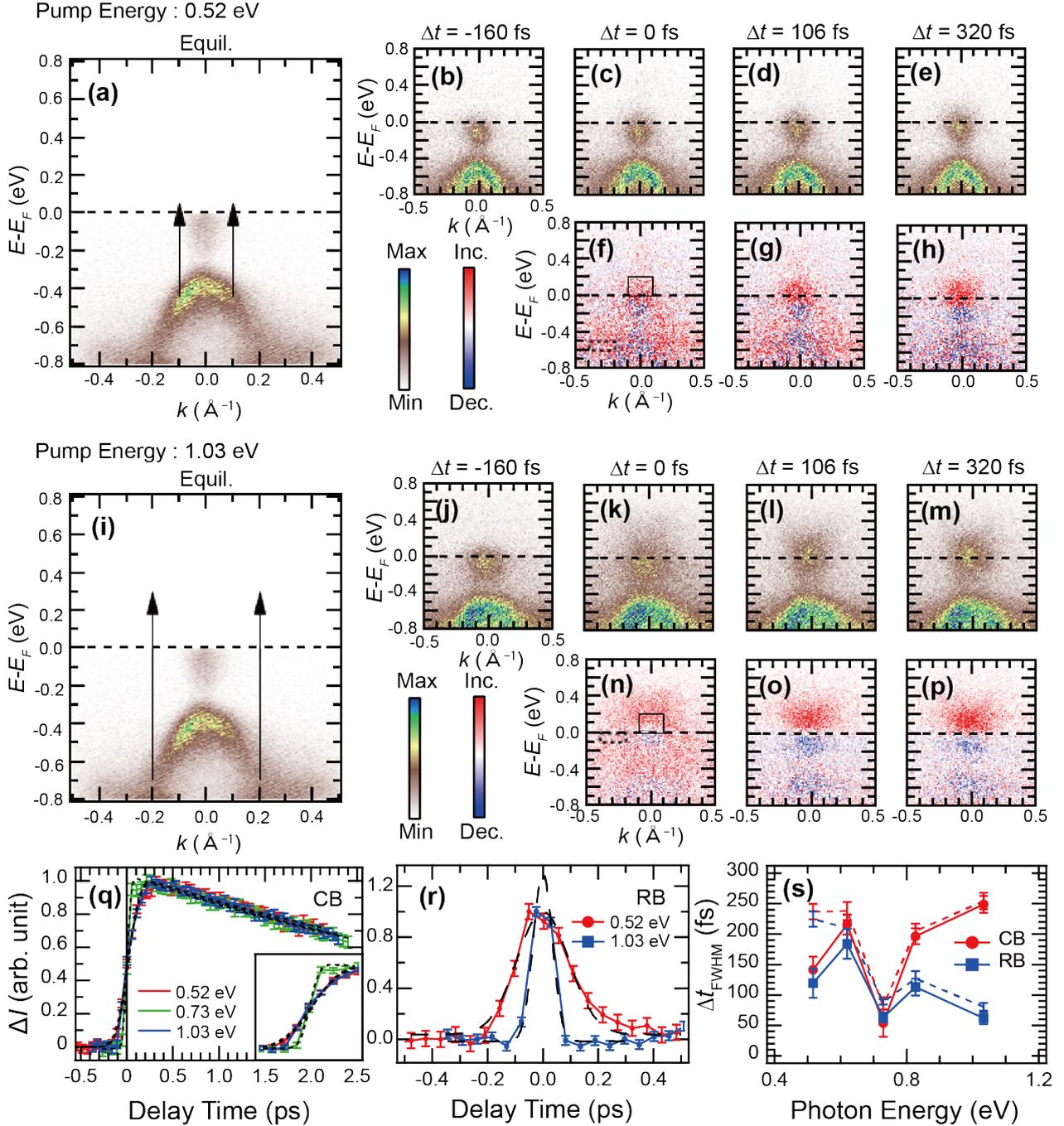

**FIG. 6** (a) Static ARPES image for Bi$_2$Se$_3$. The black arrows represent the optical excitation by the pump photon energy of 0.52 eV. (b)-(e) TARPES image for Bi$_2$Se$_3$ with the 0.53 eV pump. (f)-(h) Differential TARPES images at the corresponding delay times given in (c)-(e) obtained by subtracting (b) from each image. (i) Static ARPES image for Bi$_2$Se$_3$. The black arrows represent the optical excitation by the pump photon energy of 1.03 eV. (j)-(m) TARPES image for Bi$_2$Se$_3$ with 1.03 eV pump. (n)-(p) Differential TARPES images at the corresponding delay times given in (k)-(m) obtained by subtracting (j) from each image. (q) Time-dependent photoemission intensity at the solid-box regions in (f) and (n) corresponding to $E - E_F = [0, 0.1]$ eV and $k = [-0.1, 0.1]$ Å$^{-1}$. The result for 0.73 eV is also shown. The fits by the rise-and-decay function are shown as black dotted lines. The inset shows the magnified region of $\Delta t = [-0.2, 0.2]$ ps. (r) Time-dependent photoemission intensity at the dashed-box regions in (f) and (n) corresponding to $E = [-0.6, -0.5]$ eV and $k = [-0.4, -0.2]$ Å$^{-1}$ for 0.53 eV and $E = [-0.1, 0.0]$ eV and $k = [-0.4, -0.2]$ Å$^{-1}$ for 1.03 eV. The dashed-black lines are the fits by Gaussian. (s) Full width at the half maximum (FWHM) of the fits for the conduction band (CB) and replica band (RB) shown in (q) and (r) as a function of the pump photon energy. The total time duration is shown as dashed lines and response times using Eq. (3) are shown as solid lines.

duration of the replica bands is expected to reflect cross-correlation between pump and probe. Figure 6(r) shows the time-dependent photoemission intensity at the dashed-box in Fig. 6(f) ($E - E_F$ = [-0.6, -0.5] eV and $k$ = [-0.4, -0.2] Å$^{-1}$) and 7(n) ($E - E_F$ = [-0.1, 0.0] eV and $k$ = [-0.4, 0.2] Å$^{-1}$). The energy and momentum ranges are chosen to correspond to the region energetically above the valence bands by the pump energy. The width for 0.52 eV is larger than that for 1.03 eV. To estimate the temporal width quantitatively, we fit the spectra with Gaussians shown as the dashed lines. The temporal widths (FWHM) are shown as the blue-dashed line in Fig. 6(s) and the response time by using Eq. (3) are shown as the blue-solid line in Fig. 6(s). In principle, the time duration of the appearance of replica band corresponds to the cross-correlation between pump and probe, and response time of the replica band should be zero. However, the finite value of the replica band is clearly noticed.

As briefly described above, the nonmonotonic behavior of the response time for the conduction band with respective to the photon energy and the finite value of the replica band are unresolved observations in this work. Here, we provide possible mechanisms for each case. For the conduction band, the reason might be due to the inclusion of the relaxation process from the higher conduction band. For the lower photon energy, the electrons are directly excited to the bottom of the conduction band corresponding to the solid black boxes in Figs. 6(f) and 6(n). For the higher photon energy, on the other hand, pump excite electrons at energetically higher states in the conduction band as shown in Fig. 6(i), and the rising dynamics of $\Delta I$ include the relaxation process from the energetically higher states. The minimized value at 0.73 eV is the most unexplained behavior. One reason might be due to the resonant excitation at the higher lying conduction band, which enhance the excitation at the specific energy and momentum region and minimize the relaxation effects from the other region. Interestingly, the minimum rising time is at 0.73 eV is also observed in HOPG, in which we interpreted it due to the electron-phonon scattering. Another possibility is that there are some typical phonons around that energy region, which switch the relaxation mechanism for the electrons. For the replica band, the change of the photoemission is below $E_F$. Since the reduction of the existed electrons before pump excitation also affect the photoemission change below $E_F$, the dynamical behavior of $\Delta I$ is not so simple as that above $E_F$. The net effect between the emergence of the replica band and the reduction of the population of electrons can explain the experimental observations. To uncover these observations, more systematic measurements and investigations are necessary.

IV.    Conclusion

We have developed a HHG laser TARPES setup with wavelength-tunable excitation enabled by using twin regenerative amplifiers with a common Ti:sapphire oscillator. The spectral width of the HHG probe is 53 meV as demonstrated on polycrystalline Au films. The wavelength (photon energy) of the pump can range from 1200 to 2400 nm (0.52 to 1.03 eV) in addition to 800 nm (1.55 eV) of the fundamental light. The fluence exceeds 5 mJ/cm$^2$ at the sample position for this wavelength range. The time duration of the pump is dependent on the photon energy. The time resolution is evaluated to be better than 50 fs for the pump energy > 0.7 eV. To test the performance of the machine, we measure the two prototypical samples of HOPG and Bi$_2$Se$_3$. We also estimate the response time of HOPG and Bi$_2$Se$_3$ by subtracting the effect of the duration of the pump and probe. By employing wavelength-tunable excitation, it becomes possible to achieve resonant and non-resonant excitations in intriguing quantum materials as was observed in HOPG and Bi$_2$Se$_3$ in this work. Consequently, our developed system opens up the potential for discovering novel non-equilibrium phenomena in the future.

This work was supported by Grants-in-Aid for Scientific Research (KAKENHI) (Grant Nos. JP19H01818, JP19H00659, and JP19H00651) from the Japan Society for the Promotion of Science (JSPS), by JSPS KAKENHI on Innovative Areas "Quantum Liquid Crystals" (Grant No. JP19H05826), and the Quantum Leap Flagship Program (Q-LEAP) (Grant No. JPMXS0118068681) from the Ministry of Education, Culture, Sports, Science, and Technology (MEXT). T. S. acknowledges research grants from The Murata Science Foundation, The Hattori Hokokai Foundation, Toyota Riken Sholar Program, and Izumi Science and Technology Foundation.

AUTHOR DECLARATIONS
Conflict of interest
The authors have no conflicts to disclose.

Author contributions
**Takeshi Suzuki**: Conceptualization (lead); Data curation (lead); Formal analysis (lead); Investigation (lead); Methodology (lead); Software (lead); Writing – original draft (lead); Writing – review & editing (lead) Supervision (lead). **Yigui Zhong**: Investigation (equal); Methodology (equal); Writing – review & editing (equal). **Kecheng Liu**: Investigation (equal); Methodology (equal); Writing – review & editing (equal). **Teruto Kanai**; Investigation (equal); Methodology (equal); Writing – review & editing (equal). **Jiro Itatani**; Investigation (equal); Methodology (equal); Project administration (lead);

Writing – review & editing (equal). **Kozo Okazaki**; Conceptualization (lead); Investigation (lead); Methodology (lead); Project administration (lead); Supervision (lead); Writing – review & editing (lead).

Data availability

The data supporting the findings of this study are available from the corresponding author upon request.

*Corresponding author.
takeshi.suzuki@issp.u-tokyo.ac.jp
†Corresponding author.
okazaki@issp.u-tokyo.ac.jp